\documentclass[prd,a4paper,showpacs,nofootinbib,preprintnumbers,amsmath,amssymb]{revtex4}
\usepackage[dvipdfm]{graphicx}
\usepackage{amssymb}
\usepackage{amsmath}
\usepackage{parskip}
\usepackage{epsfig}
\usepackage{dcolumn}
\usepackage{rotating}
\newcommand{\bea}{\begin{eqnarray}}
\newcommand{\eea}{\end{eqnarray}}

\newcommand{\be}{\begin{equation}}
\newcommand{\ee}{\end{equation}}
\newcommand{\beqy}{\begin{eqnarray}}
\newcommand{\eeqy}{\end{eqnarray}}
\newcommand{\p}{\partial}

\newcommand{\mx}{\mbox}
\newcommand{\mt}{\mathtt}

\newcommand{\ga}{\gamma}
\newcommand{\Ga}{\Gamma}

\newcommand{\de}{\delta}
\newcommand{\De}{\Delta}

\newcommand{\ze}{\zeta}
\newcommand{\s}{\sigma}
\newcommand{\e}{\epsilon}
\newcommand{\om}{\omega}

\newcommand{\la}{\lambda}

\newcommand{\cO}{{\cal O}}

\newcommand{\cL}{{\cal L}}

\newcommand{\2}{\frac{1}{2}}

\newcommand{\ra}{\rightarrow}
\newcommand{\Ra}{\Rightarrow}

\newcommand{\LF}{\left(}
\newcommand{\RF}{\right)}
\newcommand{\LT}{\left[}
\newcommand{\RT}{\right]}
\newcommand{\ie}{{\it i.e.\ }}

\newcommand{\x}{\dagger}
\newcommand{\Ds}{\De^{\star}}

\begin{document}

\title{The cosmological BCS mechanism and the Big Bang Singularity}

\author{Stephon Alexander$^{ab}$}
\email{sha3@psu.edu}
\author{Tirthabir Biswas$^a$}
\email{tbiswas@gravity.psu.edu}

\vskip 3mm

\affiliation{ {\it $^a$ Department of Physics\\ Institute for Gravitation and the Cosmos\\ The
Pennsylvania State University\\ 104 Davey Lab, University Park, PA,16802, U.S.A \\} }
\affiliation{ {\it $^b$ Department of Physics and Astronomy\\ Haverford College\\ Haverford, PA,19041, U.S.A \\} }



\date{\today}

\begin{abstract}
We provide a novel mechanism that resolves the Big Bang Singularity present in FRW space-times without the need for
ghost fields.   Building on the fact that a four-fermion interaction arises in General Relativity  when fermions are
covariantly coupled, we show that at early times the decrease in scale factor enhances the correlation between pairs of
fermions.  This enhancement leads to a BCS-like condensation of the fermions and opens a gap dynamically driving the
Hubble parameter $H$ to zero and results in a non-singular bounce, at least  in some special cases.
\end{abstract}


\maketitle


\section{Introduction and Motivation}

As is well known the Standard Big Bang cosmology (SBB) successfully predicts the observed large scale expansion, thermal
properties, nucleosynthesis and the cosmic microwave background in our universe.  However, the theorems of Hawking and
Penrose prove that a curvature singularity exists in the SBB at the ``birth'' of the universe's space-time.
Cosmologists have had long held expectations that at the Planck time quantum gravitational effects might resolve the
singularity and provide a quantum bridge that connects a collapsing phase to an expanding one; otherwise known as a
bouncing cosmology.  The implementation of the bouncing scenario in quantum theories of gravity are still at the toy
model stage, since a complete description of quantum gravity is lacking.   Given this fact, we will argue that a
non-singular cosmology is possible from the effects of fermions on space-time when the scale factor starts becoming
small.

Most modern approaches towards a unified quantum theory of gravity, such as supergravity~\cite{sugra}, or
LQG~\cite{loop-review} has found the first order formalism of gravity, where one treats the vielbein and connections as
independent variables, to be the natural starting point. It is also well known that while passing from the first to the
second order formalism (by integrating out the torsion field) the covariant coupling of fermions to gravity yields a
four fermion interaction in these theories~\cite{sugra,Jacobson,alexander,perez,freidel,das}\footnote{See however~\cite{simone} for a counter example where even in the presence of the Immirzi parameter, the fermions don't change the classical equations.}.  Now, the BCS theory of
superconductivity states that if there exists a weak attractive four fermion interaction and a fermi surface, fermions
will condense into a new ground state with lower energy comprising of spin-0 Cooper Pairs.  In the context of a
homogeneous and isotropic  cosmological space-time\footnote{The interplay between  inhomogeneities in the Fermi-gas and
the metric, and especially it's implications for CMB remains an important open question.}, we will show that fermions
indeed form  Cooper pairs.   Moreover, the FLRW scale factor (or the ``volume'' of the universe) plays the role as an
effective coupling constant such that at early times the correlations between the fermions get stronger.  As a result,
we show that as we approach the singularity an energy gap opens up, contributing negative energy which, as we shall explain shortly, is one of the fundamental requirements for obtaining realistic bounces in a spatially flat or open universe. We were also able to explicitly show that at least  when the usual (positive) matter energy density has an equation of state parameter, $\om<1/3$, the negative gap energy can drive the
expansion rate of the universe to zero, leading to a bounce between the contracting and expanding phase.  For this
mechanism to work, an {\it a priori} assumption  that is required is the existence of a finite non-zero density of
the Dirac fermions (\ie we assume a particle/anti-particle asymmetry from the very beginning).  We employ robust
methods of effective field theory to demonstrate the non-singular behavior of this cosmology. We note in passing that similar ideas have been previously employed to resolve the black hole singularity and information loss problems~\cite{blackhole}.

In contrast, previous attempts to resolve Big Bang singularity (BBS) has  mostly relied on  introducing
pathological ghost fields~\cite{narlikar,freese,frampton} which either violate quantum unitarity and/or leads to
catastrophic instability~\cite{jim}. In some other bouncing and cyclic universe
scenarios~\cite{tolman,barrow,ekpyrotic}, one leaves the singularity unresolved in the hope that ``quantum gravity''
effects will eventually smooth out the singular Big Crunch/Bang transition. Although rigorous but involved no-go
theorems exist in the literature~\cite{paris-ec} as to  why it is so challenging to avoid the singularity in General
Relativity (GR),   in the context of a flat\footnote{Although our analysis is not specific to a flat universe and
readily applies to open or closed universes, for simplicity and phenomenological reasons (WMAP and other measurements
strongly constrain the flatness~\cite{wmap}) we specialize to the flat case. However, we point out that in a closed
universe setting the presence of the spatial curvature itself can allow for a resolution of the big bang
singularity~\cite{closed}, but obviously such a resolution mechanism will not work for a general manifold, such as an
open or a flat universe which makes it  somewhat less appealing.} homogeneous  isotropic cosmology it is easy to
qualitatively see this. One just has to look at the Hubble equation 
\be H^2\equiv\LF{\dot{a}\over a}\RF^2={1\over
3M_p^2}\sum_i\rho_i \ee which governs the evolution of the scale-factor $a(t)$ appearing in the
Friedman-Lemaitre-Robertson-Walker (FLRW) metric 
\be ds^2=-dt^2+a^2(t)dx_3^2\ . \ee Here, $i$ labels the different
components of matter present in the universe. Now, in the bouncing universe construction the scale factor goes through a
minimum, where $\dot{a}$, or equivalently the Hubble expansion rate, $H$, must vanish. This means that some of the
matter components must have negative energy, a property hard to realize in conventional matter. Ghost fields have
negative kinetic energies and  can mediate a bounce and therefore have been used in previous literature. One may wonder
whether ordinary scalar fields with negative potential energy can work? Unfortunately, having a negative energy
component is not sufficient for a bounce, the negative energy component has to precisely cancel  the  positive matter
component at the bounce and then redshift away faster than it, to leave the total energy density positive. In other
words one requires violation of not only the weak energy condition ($\rho< 0$) but also the null energy condition ($\rho+p< 0$)~\cite{paris-ec}. As a consequence the energy density has to increase with expansion of
the universe! It is easy to check that  ordinary scalar fields cannot achieve this.

The reason why an attractive four fermion coupling can give rise to a consistent bouncing scenario is because, firstly
the interaction energy between fermions gives rise to a negative contribution to the energy density (binding energy
between Cooper pairs). Secondly,  this energy density depends on the ``Gap'' which in turn, depends rather nontrivially
on the scale factor (or volume of the universe) via the chemical potential. It turns out that this rather non-trivial
volume dependence can violate the null energy condition temporarily as needed to have a bounce. On hind sight the fact that fermions can violate the energy conditions probably should not  come as a real surprise, even ``classically''  fermionic condensates are known to violate energy conditions and give rise to bounces~\cite{classical}.

We should point out that in the context of effective field theory there have been some successful efforts in
constructing toy models involving higher derivative actions which can evade the problem of ghosts (for instance
involving non-local modifications of gravity~\cite{BMS}, or ghost condensation mechanism~\cite{new-ekpyrotic}) but
preserve some of the useful ghost-like properties  of higher derivative theories to resolve the  BBS. (For a more
detailed review of various bouncing scenarios the readers are referred to~\cite{novello}.) However, the mechanism that
we propose is not a toy model, and has a much more generic scope as it essentially only relies on having covariant
coupling of fermions to gravity  and the presence of a  Fermi surface.

A third requirement to achieve a bounce intriguingly, turns out to be  the presence of the Immirzi parameter (associated
with Holst's generalization of the Einstein-Hilbert action) which also seems to play a key role in the nonperturbative
quantization of gravity~\cite{loop-review}. For instance,  in Loop Quantum Gravity (LQG), the quantization of the area
operator, and the existence of bouncing cosmologies~\cite{loop} depend crucially on the Immirzi parameter. What is also
suggestive is that like in LQG, we have to rely on a completely nonperturbative effect, the formation of a BCS gap to
evade the Big Bang.

This paper is organized as follows: In section II we present the first order formalism of GR and derive the four fermion
interaction.  In section III we develop the cosmological BCS theory and derive the effective potential for the
cosmological energy gap.  In section IV we analyze a special case in details to show how and when singularity
resolution can occur.  Finally we conclude and provide future outlook in section V.
\section{Four Fermion Interaction from GR}
The goal of this section is to  summarize the first order formalism of GR in the presence of fermions and derive the
four-fermion interaction necessary for BCS condensation~\cite{perez,freidel}.  Let us consider a $4$-dimensional
manifold ${\cal{M}}$ and introduce two independent fields: the tetrad, $e_m^I$, an orthonormal coordinate basis for each
point on the manifold, and a spin connection $A_{mI}{}^J$ which connects (parallel transports) the tangent spaces at
different points of the manifold. Note that small-case Latin letters starting from $m,n\dots$ denote spacetime indices,
while capital Latin letters starting from $I,J\dots$ denote internal Lorentz indices. One can then associate a
$4$-dimensional metric $g_{mn}$ via
\be g_{mn} = e_m^I e_n^J \eta_{IJ}, \ee

where the Minkowski metric is to be
viewed as the metric of the internal space. Internal indices are raised and lowered with the Minkowski metric, while
spacetime indices are raised and lowered with the spacetime metric. Using the connections one can define the covariant
derivatives of mixed tensors via usual rules such as
\be D_{m} k_{nI} = \partial_m k_{n I} + A_{mn}{}^{p} k_{p I} +A_{m I}{}^J k_{n J},\mx{ where }A_{mn}{}^{p}\equiv e_n^I e_J^pA_{mI}{}^J \ee
 $D_m$ is known as the  generalized covariant derivative. As is evident, we use the tetrad and its inverse, $e_J^p$ to
 convert internal indices to space-time indices and vice-versa. The requirement
that the spin connection be torsion free is simply $A_{[mn]}{}^p = 0$.

At this juncture we should briefly discuss the relation between internal and spacetime quantities. Riemannian fields,
like the metric or the Einstein tensor, live on some substructure of the {\emph{base
    manifold}} ${\cal{M}}$, such as the tangent or cotangent spaces,
and thus, have a fixed dimension (that of ${\cal{M}}$). Alternatively, gauge fields live in an {\emph{internal}} vector
space, which is independent of the base manifold and could in principle be infinite dimensional. A {\emph{fiber bundle}}
is then simply the union of the base manifold and the internal space, where each fiber is a copy of the vector space
corresponding to a particular space-time point. {\it Such fiber bundles have an associated structure group or Lie group,
which qualitatively glues all fibers together.} In the context of the first order formalism it is convenient to view
$m,n\dots$ as associated with the cotangent/tangent spaces, while  $I,J\dots$  with  internal vector indices associated
with the Lie Group $SO(3,1)$.

With the generalized covariant derivative and spin connections we can now define the generalized curvature tensors. This
is done through the failure of commutativity of the generalized covariant derivatives. One defines
\begin{subequations}
\bea R_{mnI}{}^J &=& \partial_{[m} A_{n] I}{}^J + \left[A_m,A_n\right]_I{}^J,
\\
\Ra R_{mnp}{}^q &=&  \partial_{[m} A_{n] p}{}^q + \left[A_m,A_n\right]_p{}^q, \eea
\end{subequations}
where the commutator is short-hand for
\be \left[A_m,A_n\right]_I{}^J := A_{mI}{}^K A_{nK}{}^{J} - A_{nI}{}^K
A_{mK}{}^{J}. \ee
Note that if the connection is metric compatible and torsion-free (ie, if it is the Christoffel
connection), then the curvature tensor is simply the Riemann tensor.

Let us now rewrite the Einstein-Hilbert action in terms of these new variables. Note, however, that we wish to work with
the trace of the generalized curvature tensor, and not the Ricci scalar, since these two quantities are not necessarily
equivalent. The Einstein-Hilbert action is given by the well-known expression
\be \label{S_E} S_{E} =\frac{M_p^2}{2}
\int d^4x\ \sqrt{-g}R= \frac{M_p^2}{2} \int d^4x \;\;  ee^m_I e^n_J P_E^{IJ}{}_{KL}R_{mn}{}^{KL},\mx{ with
}P_{E}^{IJ}{}_{KL}\equiv \de_K^{[I}\de_L^{J]}
\ee
where $E$ simply stands for ``Einstein-Hilbert''.
 The second relation can be derived  using the
identity
\be R = \2\delta^{m}_{[p} \delta^n_{q]} R_{mn}{}^{pq}, \ee

We note in passing that in the absence of fermions, the field equations for the connection gives rise to the
compatibility (or zero torsion) condition: \be D_{[m}e_{n]}^a=0 \label{compatibility} \ee which means that $A_{mI}{}^J$
are nothing but the Christoffel symbols determined in terms of the metric and one recovers Einstein's GR. This statement
remains true even when one includes Holst's modification to (\ref{S_E}): \be P_{E}^{IJ}{}_{KL}\longrightarrow
P_H^{IJ}{}_{KL}\equiv \de_K^{[I}\de_L^{J]}-{1\over2\ga}\e^{IJ}{}_{KL} \ee One can check that once one imposes the
compatibility condition (\ref{compatibility}), the additional term in Holst's action, $S_H$ (which has $P_H$ rather that
$P_{E}$ in (\ref{S_E})), vanishes and therefore does not effect the classical equations of
motion.

However, things dramatically change when one includes  the covariant coupling of free fermions to GR in the presence of
the Holst term:
\begin{subequations}
\bea S &=& S_{H} + S_{D}
\\
\label{Dirac-S} S_{D} &=&- \frac{1}{2} \int d^4x \sqrt{-g} \left(i \bar{\psi} \gamma^I
  e_I^m {\cal{D}}_m \psi + {\textrm{c.c.}} \right),
\eea
\end{subequations}
$S_D$ corresponds to the Dirac action with massless fermions~\footnote{The inclusion of a mass term does not affect the conclusions of this
paper, but we leave it out to avoid cluttering.}, where ${\textrm{c.c.}}$ stands for complex conjugation, $\psi$ is a
Dirac spinor, and $\gamma^I$ are $4\times 4$ gamma matrices defined via \be \gamma_I=\left(
\begin{array}{cc}
0&\s_I\\ \bar{\s}_I&0
\end{array}
\right) \mx{ with }\s_I\equiv(1,-\vec{\s})\mx{ and }\bar{\s}_I\equiv(1,\vec{\s}) \ee where $\vec{\s}$ are just the usual
2 by 2 sigma matrices. With these definitions it is easy to check that the gamma matrices  obey the following
anticommuting algebra \be \{\ga_I,\ga_J\}=-2\eta_{IJ} \ee In (\ref{Dirac-S}) we have also defined \be
\bar{\psi}=\psi^{\dagger}\ga_0=\psi^{\dagger} \left(
\begin{array}{cc}
0&1\\ 1&0
\end{array}
\right) \ee

Note that a tetrad based formalism is essential for the inclusion of fermions in the theory, since Dirac spinors live
naturally in $SU(2)$. Therefore, covariant derivative associated with the Dirac action are not the usual $SO(3,1)$
covariant derivatives, but instead are given by ${\cal{D}}_m \psi := \partial_m \psi - (1/4) A_m{}^{IJ} \gamma_I
\gamma_J \; \psi$.

Let us now find the structure equations of the fermion-extended theory. In general one can break up the connection into
symmetric and antisymmetric pieces: \be A_m^{IJ}=\om_m^{IJ}+C_m^{IJ} \ee where $\om_m^{IJ}$ is the torsion-free spin
connection satisfying the compatibility condition (\ref{compatibility}) and $C_m^{IJ}$ is the so called ``contorsion''
tensor. The idea is to integrate out the the contorsion tensor which then will lead us to the more familiar second order
formulation of gravity where the connection are just the metric dependent Christoffel symbols. This can be achieved
simply by imposing the structure equations obtained by varying the action with respect to the connection.
Following~\cite{perez} and using the identity \be \gamma^{I} \gamma^{[J} \gamma^{K]} = - i \epsilon^{IJKL} \gamma_{5}
\gamma_{L} + 2 \eta^{I[J} \gamma^{K]}, \ee we can express the contorsion tensor in terms of the axial fermion
current,$J_5^I := \bar{\psi} \gamma_5 \gamma^I \psi$ \be e_I^mC_{mJK}=4\pi G {\ga^2\over \ga^2+1}\LF
\2\e_{IJKL}J_5^L-{1\over \ga}\eta_{I[J}J_{5K]}\RF \label{contorsion} \ee From the above expression for the contorsion
tensor it is clear that $C_{mJK}$ is a non-propagating field, its field equations do not have any derivatives on it.
Thus ``integrating it out'' is not only equivalent to reinserting its expression (\ref{contorsion})  in the full action
classically, but also quantum mechanically. Thus the four-fermion contact interaction term that we are going to generate
in going from the first to second order formalism is quantum mechanically an exact result. This is a key difference from
the ``effective'' contact interaction that one obtains in non-abelian gauge theory where the mediating gauge fields do
indeed propagate, and therefore the contact term is only a low energy approximation.

Substituting (\ref{contorsion}) in Holst's action we  find that the action can be written as~\cite{perez} \be
\label{split-action} S = S_{E}[\omega] +  S_{D}[\omega] + S_{int}. \ee

The first and the second terms are the standard Einstein-Hilbert and Dirac action involving the Christoffel connections.
However, crucially one obtains a  third  interaction term, given by\footnote{The sign difference in front of the
four-fermion term as compared to what was derived in \cite{perez,freidel} is simply because our metric has the opposite
signature.} \be \label{inter-term} S_{int} =  \frac{3}{2}\pi G \LF{\ga^2\over \ga^2+1}\RF\int d^{4}x \; e \;J_{5I}
J^I_5 \equiv \int d^{4}x \; e \;{J_{5I} J^I_5\over M^2} \ee Such four-fermion interactions were already observed in
Einstein-Cartan theory ($\ga^2\ra\infty$ limit in (\ref{inter-term})), although they are suppressed by a power of
Newton's constant (a factor of $1/\kappa$ here).  In order to form a condensate of fermions the Planck suppression will
have to be transcended, and this happens when the universe contracts to Planck densities, as we shall discover in the
next section. 
\section{Cosmological BCS theory}
In this section the starting point of  our discussion is the General Relativistic action (\ref{split-action}) derived in
the previous section, but we want to study it in a condensed matter framework, \ie not in vacuum but in the presence of
a gas\footnote{Perhaps it is better to call the system as a Fermi-liquid~\cite{fermi-liquid} rather than a Fermi gas, as
they are interacting. Also, in principle there could be several species of fermions, but for simplicity we will only
consider one.} of  fermions. In this case, one has to add the contribution of the chemical potential ($\mu$) to the
effective action, which at zero temperature reads~\cite{qcd,bcs-bec}:
\be S=\int d^4x e^{-1}\LT {M_p^2\over
2}R-i\bar{\psi}\ga^ID_I\psi -\mu\bar{\psi}\ga^0\psi+{J_{5I} J^I_5\over M^2}\RT \label{action} \ee
We note in passing
that although here we consider four fermion coupling  arising from covariant coupling of fermions to gravity via the
torsion constraint,  an ``effective'' gauge mediated contact interaction term~\cite{qcd} can also lead to similar
mechanisms, and could be interesting to pursue in the future.

Our aim is to study in details the  possibility of a  cosmological BCS-like condensation of the fermions first suggested
in~\cite{deepak} and to consider its implications  towards the resolution of the Big Bang singularity. As discussed in
the introduction,  to have  a resolution of BBS, we have to ensure  that the Hubble expansion rate can vanish. As a
first approximation then it becomes sufficient to look at the condensation mechanism on a flat Minkowski background \ie
ignoring the space-time dynamics; one can readily convince oneself that corrections to due to expansion of the universe
are expected to be $\cO(H/\mu)$ and therefore are suppressed near a possible bounce  when $H\ra 0$.

Let us therefore focus on the Minkowski space-time action: \be S=\int d^4x \LT-i\bar{\psi}\ga^m\p_m\psi
-\mu\bar{\psi}\ga_0\psi+{J_{5m} J^m_5\over M^2}\RT \label{action} \ee Technically it is simpler to work in terms of two
component Weyl fermions, $\ze_{F,A}$, where  the left handed fermion sector is denoted by $F$ (it contains the left
handed fermion and right handed anti-fermion) and the left handed anti-fermion sector is denoted by $A$ (it contains the
left handed anti-fermion and right handed fermion). Let us first look at the free part of the action: \be S=\int d^4x
\LT i(\ze^{F\x}\bar{\s}^m\p_m\ze^{F}+\ze^{A\x}\bar{\s}^m\p_m\ze^{A}) -\mu(\ze^{F\x}\ze^{F}-\ze^{A\x}\ze^{A})\RT
\label{weyl-action} \ee

A simple and physically transparent way to understand the condensation mechanism is to  introduce auxiliary scalar (gap)
fields,  which are proportional to the fermionic bilinears. The so called ``gap equation'' is then derived by
integrating out the fundamental fermionic degrees of freedom. As is well-known~\cite{qcd,bcs-bec}, in these integrals
the contribution of the gap is inversely proportional to the energy of the fermions. Now each of the  $\ze$'s describe
two physics degrees of freedom, corresponding to the positive and negative helicity states $-i\vec{\p}.\vec{\s}\
\zeta=\vec{\s}.\vec{p}\ \zeta=\pm| \vec{p}|\ \zeta$. Thus we see that the positive helicity state for  $\ze^F$ and
the negative helicity state  for $\ze^A$ (\ie the left and right handed fermions) correspond to energies  $E= \pm||
\vec{p}|-\mu|$ (the positive  and the negative sign corresponds to  ``particle'' or ``hole'' like excitations around the
Fermi surface respectively). Thus  close to the Fermi surface the energy of these fermionic states can vanish and
thereby contribute  significantly to the gap energy. On the other hand, the antifermionic states corresponding to
negative helicity and positive helicity states for  $\ze^F$ and   $\ze^A$ respectively, always have non-vanishing
energies $\sim | \vec{p}|+\mu$ and hence their contribution towards the gap is suppressed (see for instance~\cite{qcd}
for a discussion on this point). In the following analysis therefore we are only going to focus our attention on the
fermionic states  and couplings between them.

A simple way to  project out the anti-particle states, as advocated in~\cite{qcd}, is to first go into the momentum
basis: \be \zeta(x)={\int d^4p\over (2\pi)^4} \ze_p e^{ipx}\mx{ where }\zeta_p=\int d^4x\ \ze(x) e^{-ipx}\ . \ee Then
for a given momentum mode choose a reference frame where the momentum is aligned along the positive $z$-axis to compute
the different terms in the action, and  covariantize the final result. It is easy to see that the free action becomes
\be \int d^4x\cL_{\mt{fer}}=\int {d^4p\over (2\pi)^4} \LT\zeta^{L\dagger}_{p}(p^0+\e_p)\zeta_p^L
+\zeta^{R\dagger}_{p}(p^0-\e_p)\zeta_{p}^R\RT\mx{ where }\e_p\equiv |\vec{p}|-\mu \ee
where $\ze^L$ and  $\ze^R$ are now
single component fermions corresponding to the left and right handed fermionic states. When $\vec{p}$ is along the
direction of the positive (negative) $z$-axis they are the upper (lower) and lower (upper) components of  $\ze^F$ and
$\ze^A$ respectively.

The four fermion term simplifies as the following: Firstly
$$J^{5m}J^{5}_{m}=\eta_{mn}(\ze^{F\x}\bar{\s}^{m}\ze^F+\ze^{A\x}\bar{\s}^{m}\ze^A)(\ze^{F\x}\bar{\s}^{n}\ze^F+\ze^{A\x}\bar{\s}^{n}\ze^A)=2\eta_{mn}\ze^{F\x}\bar{\s}^{m}\ze^F\ze^{A\x}\bar{\s}^{n}\ze^A$$
The second equality stems from the fact that we are only interested in coupling between $\ze^F$ and  $\ze^A$, or more
precisely between $\ze^L$ and  $\ze^R$ (all other terms when projected onto fermionic states will be of the form
$(\ze^{L}\ze^{\x L})^2$ or $(\ze^{R}\ze^{\x R})^2$ which must vanish due to anti-commutavity). This further enables us
to consider only the diagonal $\bar{\s}^{m}$ matrices (the off-diagonal $\bar{\s}^{m}$'s only contain couplings between
fermionic and anti-fermionic states):
$$J^{5m}J^{5}_{m}=2[-\ze^{F\x}\bar{\s}^{0}\ze^F\ze^{A\x}\bar{\s}^{0}\ze^A+\ze^{F\x}\bar{\s}^{3}\ze^F\ze^{A\x}\bar{\s}^{3}\ze^A+\dots]=-4\ze^{L\x}\ze^L\ze^{R\x}\ze^R=4(\ze^{L\x}\ze^R)(\ze^{L\x}\ze^R)^{\x}$$
 Having found the desired left-right coupling, one can now introduce the auxiliary fields $(\De,\Ds)$, so
that we can rewrite the four fermion coupling as \be 4\int d^4x{(\ze^{L\x}\ze^R)(\ze^{L\x}\ze^R)^{\x}\over M^2}=\int
d^4x \ \LT\Ds\ze^{L\x}\ze^R+\De\ze^{R\x}\ze^{L}-{M^2\over 4}\De\Ds\RT \ee It is clear that the auxiliary field, $\De\sim
\ze^{L\x}\ze^R$, and a non-zero value for it would signal a cosmological BCS-like condensation! In order to find such a
nontrivial value for $\De$, one can now takes recourse to a mean-field approximation where $\De$ is  treated as a
constant ``gap''.

To obtain the ``gap equation'' (or the effective theory of $\De$) we  have to integrate out the fundamental degrees of
freedom, the fermions. In the momentum basis the fermionic action looks like \be \int d^4x\cL_{\mt{fer}}=\int {d^4p\over
(2\pi)^4} \LT\zeta^{L\dagger}_{p}(p^0+ \e_p)\zeta_p^L
+\zeta^{R\dagger}_{p}(p^0-\e_p)\zeta_{p}^R+\Ds\ze^{L\x}\ze^R+\De\ze^{R\x}\ze^{L}\RT \ee Or, in matrix notation \be \int
d^4x\ \cL_{\mt{fer}}=\int {d^4p\over (2\pi)^4}\ (\zeta^{L\dagger}_{p},\ze^{R\x}_{p}) \LF\begin{array}{cc} p^0+ \e_p&
\De\\ \Ds& p^0-\e_p
\end{array}\RF
\LF
\begin{array}{c}
\zeta^{L}_p\\ \ze^{R}_{p}
\end{array}
\RF\equiv \int d^4p\ (\zeta^{L\dagger}_{p},\ze^{R\x}_{p})A_p \LF
\begin{array}{c}
\zeta^L_p\\ \ze^{R}_{p}
\end{array}
\RF \ee

As usual in the path integral when one integrates over all the fermionic variables $\xi_p$'s, one ends up with a
fermionic determinant. In general for a fermionic path integral we have (see~\cite{qcd,bcs-bec}, for instance) \be
Z\equiv \int {\cal D}\xi e^{iS}=\int {\cal D}\xi \exp\LT i\int d^4p\ \xi^{\dagger}_pM_p \ze_p\RT=\exp\int {d^4p\over
(2\pi)^4}\ tr(\ln M_p) \ee where $\xi$ in general can be a $N$ component spinor and $M_p$ is an $N$ by $N$ matrix. The
effective action $S_{\mt{eff}}$ (in Quantum Field Theory literature this is often reffered to as  $\Ga_{\mt{eff}}$) is
defined in terms of the path integral via \be Z\equiv e^{iS_{\mt{eff}}} \ee Thus for our path-integral we obtain the
following effective Lagrangian for $\De$ (for a more detailed derivation see~\cite{bcs-bec}) \be
\cL_{\mt{eff}}(\De)=-i\int {d^4p\over (2\pi)^4}\ tr(\ln A_p)-{M^2\over 4}\De\Ds \ee
The above effective action can
actually be evaluated exactly. Focussing on the first (non-perturbative) term which was obtained by integrating out the
fermions   we find
\be \cL_{\mt{non}}(\De)\equiv\int {d^3p\over (2\pi)^3}\sqrt{\e_p^2+\De^2}=\int_{-\la}^{\la} {d\e_p\
(\e_p+\mu)^2\over 2\pi^2}[ \sqrt{\e_p^2+\De^2}] \ee
where in the last step we have performed the angular integrations
and we are considering modes (electrons or holes with positive or negative $\e_p$ respectively) around  $\mu$, $\la$ is the UV cutoff used to regulate the integrals. The above
integral is formally divergent. However, employing the above cut-off regularization scheme we first of all can
subtract the quartic divergence by requiring $\cL_{\mt{non}}(\De)$ to vanish when $\De=0$, or in other words just
subtract the  $\De=0$ piece from  $\cL_{\mt{non}}(\De)$\footnote{The quartically divergent piece is nothing but the contribution of the fermions  to the vacuum energy. This is  a specific illustration of the famous ``cosmological constant problem'': Quantum loop contributions to the cosmological constant are generally known to go as $\sim \cO(M_p^4)$ for a theory with a Planck cut-off scale, but observations suggest that we are living in an universe with a much smaller cosmological constant $\sim 10^{-120}M_p^4$. This implies an incredible cancelation of a hundred and twenty orders of magnitude between the different contributions to the vacuum energy. Since in this paper we are not trying to address the cosmological constant problem, we take the usual approach and assume that some unknown mechanism is indeed responsible for such a cancelation/``renormalization'' of the vacuum energy down to it's extremely small observed value, $\sim \cO(mev^4)$. Since the bounce in our model is governed by much higher energy scales $\sim M_p^4$, such a tiny  cosmological constant will play no role. If on the other hand the vacuum energy at the bounce was large and only became small later, it can modify the nature of the bounce. However as we shall see later, whether we  have a bounce or not only depends on the equation of state of matter, $\om$, and in particular as long as $\om<1/3$ the Big Crunch/Bang singularity is resolved via the bounce. Since the cosmological constant has an equation of state $\om=-1$, it's existence does not pose a problem for our mechanism.}
Unfortunately, a logarithmic divergence remains in the form of an undetermined mass parameter $\e_r\ (=2\la)$ due to the presence of the four fermion interaction,  an artifact of the  intrinsically non-renormalizable torsion-gravity theory. Even on dimensional grounds it is
clear that such a cut-off must be present. In condensed matter systems, this cut-off corresponds to the Debye frequency
related to the lattice spacing. It seems likely that in case of gravity  $\e_r$ would similarly be related to the
fundamental quantum (discretization) of space-time as for instance observed in Loop Quantum gravity~\cite{loop-review}.

Putting everything together therefore we have the final effective potential  given by
\be
V_{\mt{tot}}=-\cL_{\mt{eff}}(\De)={M_r^2\De^2\over 4}-{1\over 2\pi^2}\LT{\De^4\over
16}+{\mu^2\De^2\over2}+\De^2\LF{\De^2\over 4}-\mu^2\RF\ln{\De\over\e_r}\RT \ee
This has one arbitrary parameter $\e_r$
as promised, along with the inverse coupling constant $M_r$. It is clear that the above potential has a minimum at say
$\De=\De_0$ given by
\be {\p V_{\mt{tot}}\over \p \Delta}=0\Ra  M_r^2={\De_0^2\over 2\pi^2}+\LF{\De_0^2-2\mu^2\over
\pi^2}\RF\ln{\De_0\over \e_r} \label{gap} \ee
One can think of the above equation as specifying $\De_0=\De_0(\mu)$. It
is useful to check that we recover the usual behavior of the Gap in the weak coupling BCS limit~\cite{qcd,bcs-bec} when
the fermion gas is dilute.   For $\De\ll\mu,\e_r$, (\ref{gap}) tells us
\be \De\approx \e_r\exp\LF{-M_r^2\pi^2\over
2\mu^2}\RF \label{exp-gap} \ee
 the familiar exponentially suppression of the gap appears.

Returning now to the general discussion, we can also calculate
\be V_{\mt{min}}\equiv V_{\mt{tot}}(\De_0)={\De_0^2\over
4\pi^2}\LT{3\De_0^2\over 8}+{\De_0^2\over 2}\ln{\De_0\over \e_r}-\mu^2\RT
\ee
However, the potential energy that we have
calculated includes the contribution from the chemical potential as well. This is the ``extra'' $\mu$ dependent term in the Lagrangian (\ref{action}); note that the number density, $n$, of fermions is given by~\cite{bcs-bec}
\be
n=\int d^4 x \ e^{-1}\bar{\psi}\ga^0\psi={\p S\over\p \mu}=- {\p V_{\mt{tot}}\over \p \mu}
\label{n-mu-relation}
\ee

In order to do cosmology what we really need
is the total energy density which is therefore related to $V_{\mt{tot}}$ via
\be \rho_{\mt{gap}}=V_{\mt{tot}}+\mu n \ee
In particular we need to know the dependence of the gap energy density as a function of the scale factor/number density. This is provided by the implicit relation between the number density and the chemical potential (\ref{n-mu-relation}):
\be n=-{\p V_{\mt{tot}}\over \p \mu}={\mu\De_0^2\over
2\pi^2}\LF1-2\ln{\De_0\over \e_r}\RF \label{number-density}
\ee
Thus putting everything together we have
\be
\rho_{\mt{gap}}=V_{\mt{tot}}-\mu {\p V_{\mt{tot}}\over \p \mu}={\De_0^2\over 4\pi^2}\LT{3\De_0^2\over
8}+\mu^2+\LF{\De_0^2\over 2}-4\mu^2\RF\ln{\De_0\over \e_r}\RT \label{energy} \ee
Using (\ref{gap}), (\ref{number-density}) and (\ref{energy}) one in principle knows the dependence of the energy density as a function of the fermion number density, $\rho_{\mt{gap}}=\rho_{\mt{gap}}(n)$.

\section{BEC and Bounce}
For the present paper, we are interested in homogenous and isotropic FLRW type cosmology  focussing on the regime
$H\ll\mu$, relevant for discussing nonsingular cosmological models. In particular what we would like to demonstrate is that in the presence of a fermion condensate,  the universe does not contract all the way to a singularity where the energy densities go to infinity, but rather bounces back at some finite energy density to an expanding phase thereby resolving the singularity. Now, in a realistic universe the  matter content of the universe will be varied, but for the purpose of illustration here we will assume that the matter content of the universe, apart from the condensate,  is described by an ideal fluid, $\rho_{\mt{mat}}$, with an equation of state parameter $\om$:
 \be
 \Ra \rho_{\mt{mat}}=\rho_0 a^{-3(1+\om)}
 \label{rho-matter}
 \ee
To check the robustness of the bounce mechanism  we should check whether or not the universe can bounce for any equation of state parameter obeying the weak energy condition, \ie $-1\leq \om\leq 1$.  In other words, we want to see when the Hubble equation
\be
H^2={1\over 3M_p^2}(\rho_{\mt{gap}}+\rho_{\mt{mat}})\equiv {\rho_{\mt{tot}}\over 3M_p^2}
\ee
allows for a bouncing solution. In order to solve the above equation we need to know how the energy densities evolve as a function of the scale factor. For matter, since it is treated as an ideal gas this is given by (\ref{rho-matter}).
For $\rho_{\mt{gap}}$, one does not have an explicit expression in terms of the scale factor, but it is implicitly defined in the following way: First of all we know that the number density of the fermions scale as inverse volume:
\be
n=n_0 a^{-3}
\label{number-scale}
\ee
Now, (\ref{gap}) implicitly determines $\De_0=\De_0(\mu)$, so that (\ref{number-density}) can  be thought of as relating the number density and the chemical potential. In other words, (\ref{gap}) and (\ref{number-density}) let's us determine the chemical potential and the gap as a function of the number density. Since the energy density is given in terms of the gap and the chemical potential,  (\ref{energy}) implicitly determines the energy density in terms of the number density.   Finally using (\ref{number-scale}) we, in principle can determine the energy density as a function of  the scale factor.

In order to demonstrate that the universe indeed bounces, we have to show two things: (i) the total energy density must vanish, $\rho_{\mt{tot}}=0$, at some scale factor, say $a=a_b$.(ii) Also, $\ddot{a}>0$ at $a=a_b$ to ensure that we indeed have a minimum (and not a maximum as happens in a turnaround) for the scale factor. Differentiating the Hubble equation we find that the latter condition is equivalent to showing that $d\rho/da$ is positive at the bounce point, in violation with the null energy condition. In principle, one can scan the entire parameter space\footnote{Actually there is a gauge redundancy in the set of parameters since one can always re-scale the scale-factor.} $\{\rho_0,\om,n_0,\e_r,M_r\}$ numerically to determine when the singularity is resolved, but this a rather involved and challenging task which we leave for future. In this manuscript, we want to consider a particular limit where the expressions simplify considerably. Let us consider the strong coupling BEC (Bose Einstein Condensation) limit~\cite{bcs-bec}, $M_r\ra0$, and when $\De\gg \e_r$. In this case (\ref{gap}) simplifies to give us a rather simple relation between the chemical potential and the gap:
\be
\De_0^2\approx 2\mu^2
\label{gap-sim} \ee
The expressions for the energy and the number density also simplifies considerably:
\bea
n=n_0 a^{-3}&\approx& {\De_0^3\over
\sqrt{2}\pi^2}\ln{\De_0\over \e_r}\Ra \De_0\sim a^{-1}\label{delta-special}\\
\mx{and }\ \rho_{\mt{gap}}&\approx &-{3\De_0^4\over 8\pi^2}\ln{\De_0\over \e_r}\sim -a^{-4}\label{rho-special}
\eea
where in inferring the scale-factor dependence of the gap and the energy density we have ignored the logarithms. Thus approximately, the gap energy density behaves as radiation with a negative sign. It is worth noting that Casimir energy calculations are also known to lead to similar negative radiation like behavior for  massless minimally coupled fields~\cite{casimir}, as the case we are considering.

\begin{center}
\begin{figure}
\includegraphics[height=6cm,angle=-0,scale=1.5]{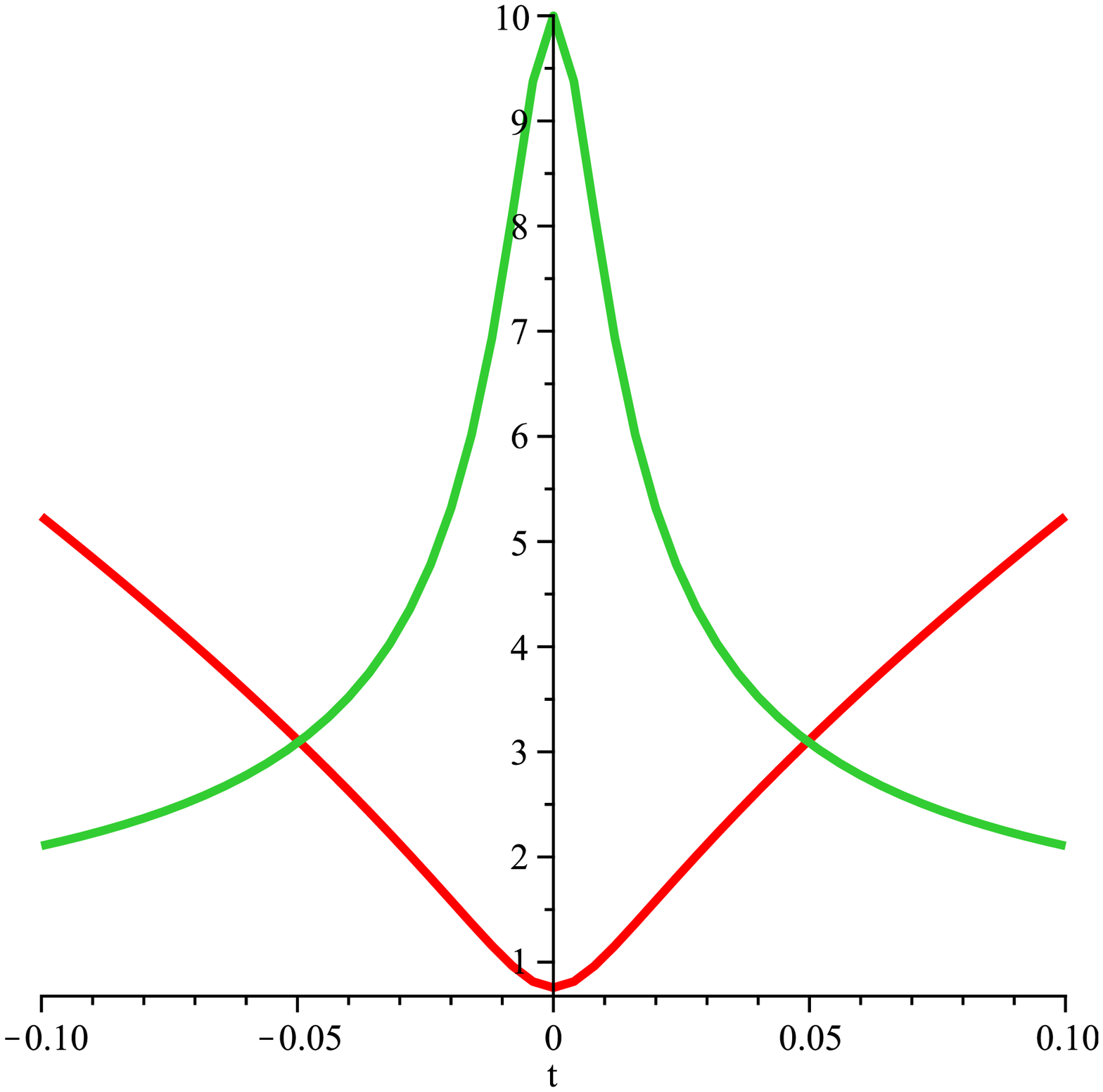}
\begin{caption}
{\small Plot of the scale factor (red curve) and the gap (green curve) as a function of time as the universe bounces. This plot corresponds to $\om=0$, and $\e_r,M_r\ll \De_0$. }
\end{caption}
\label{fig:1}
\end{figure}
\end{center}

What the above behavior suggests is that as long as $\om<1/3$, \ie the matter energy density blue shifts (during contraction) slower than radiation, we will have a bounce, see figure \ref{fig:1}. Since $\rho_{\mt{tot}}$ is always positive, the matter energy density will always dominate over the gap energy density, but during contraction since the gap energy is blue shifting faster as compared to matter, it will eventually catch up with matter and precisely cancel it at the bounce point. After the bounce, in the expanding phase the gap energy will dilute faster than matter ensuring that   $\rho_{\mt{tot}}$ remains positive. In our case, one finds that for
\be
\rho_{\mt{tot}}\approx \rho_{\mt{bounce}}
\LT \LF{a\over a_b}\RF^{-3(1+\om)}-\LF{a\over a_b}\RF^{-4}\RT\mx{ we have } \left.{d\rho\over da}\right|_{a=a_b}=1-3\om
\label{bounce}
\ee
and thus as argued before we satisfy the bounce criteria when $\om<1/3$.

A few remarks are now in order. Firstly we realize that in the strong coupling BEC limit that we are considering here, whether we will have a bounce or not only depends on $\om$. In general however,  the dependence will be much more complicated and only future numerical exploration will be able to provide a comprehensive answer. Let us next look at the robustness of the mechanism in resolving the Big crunch/bang singularity. Usually it is believed that in a contracting universe at high energy densities the universe will come to be dominated by radiation which goes as $a^{-4}$ and has an equation of state $\om=1/3$. Whether we can have a bounce, or not, in this limiting case of $\om=1/3$ crucially depends on the log like behavior of the gap energy. By inspection  one finds that because of the presence of the logarithm
in (\ref{delta-special}), $\De_0$ blue shifts slightly slower than $1/a$, and  again by inspection from (\ref{rho-special}) one deduces that $\rho_{\mt{gap}}\sim n \De\sim a^{-3}\De$, and therefore blueshifts slightly slower than radiation. This means that unfortunately in the presence of pure radiation (if other forms of matter are present, the situation may change), one cannot bounce back. Although this is  a  drawback of the proposed mechanism, according to string theory this may not necessarily be a  serious problem. According to string theory at high enough energy densities, close to the string scale, the thermal state of the universe is no longer describable by massless radiation modes. Rather, one enters a stringy ``Hagedorn phase'' where even the massive modes are excited and are in thermal equilibrium~\cite{hagedorn}. In this phase, the total matter energy density (now consisting of both massless and massive modes) actually blue(red)shifts as $a^{-3}$, and therefore the presence of condensates can indeed produce a bounce.

One may also be worried about the problem of Mixmaster chaotic behavior near the bounce, as the anisotropies are known to grow as $a^{-6}$ as the universe contracts. Since $\rho_{\mt{gap}}$ blueshifts slower than anisotropies, our bounce mechanism is indeed susceptible  to this problem. However, recent work on cyclic cosmologies involving many bounces have been shown to, at least, ameliorate the problem~\cite{emergent}.

Finally, what about the maximal energy density at the bounce point, what parameters does it depend on? As remarked earlier, since one can arbitrarily  re-scale  the scale factor, only a combination of the parameters $n_0,\rho_0$ is physical. In the special case when $\om=0$ (as one expects in the Hagedorn phase) for the limiting regime under consideration, one can check that
\be
\rho_{\mt{bounce}}={256\over 27\pi^2}{\rho_{\mt{mat}}^4(t)\over n^3(t)}={256\over 27\pi^2}{\rho_0^4\over n_0^3}
\ee

To summarize, in our picture the universe starts out with a dilute gas of fermions with an exponentially suppressed gap
(\ref{exp-gap}). This would  be completely overwhelmed by finite temperature effects and we will just have a theory of
ordinary non-interacting fermions. However, as the universe contracts and the number density $n$  increases, so does the gap energy. (Increasing $\mu$ is equivalent to decreasing $M_r$, or increasing the coupling
of the four fermion interaction. In this sense the volume or the scale-factor controls the strength of the interaction.)
Eventually at extreme high energy densities Cooper pairs are formed (superconducting phase), and the negative
interaction energy starts to cancel the positive kinetic energy contributions more and more. Finally when $\De\gtrsim
\e_r$, we expect these Cooper pairs to condense more and more in the new ground state and  the interaction energy can completely cancel the matter energy density leading to the bounce.

\section{Conclusion}
In this paper we have presented a novel physical mechanism which self consistently resolves the initial big bang singularity. The crucial ingredient relies on the transient violation of the null
energy condition during the bounce.  The interactions responsible for the negative gap energy at the bounce is enhanced
as the universe contracts to the singularity.  Typically it is difficult to obtain a bounce without the introduction of
dangerous ghost states, but in this mechanism the negative energy arises from the binding energy associated with the
bound state of the Cooper pair.   Moreover, the gap only becomes significant at early times and consistently redshifts
away at late times, in other words the late time cosmology is protected from the negative energy contribution that
generated the bounce. In a special region of the parameter space we were explicitly able to obtain bouncing solutions in the presence of matter provided it's equation of state parameter was less than one-third.

While our results are promising, a few developments are in order.  First, we have not included perturbations of the
condensate which is expected to affect the homogeneity and isotropy both at early and late times; how robust is the BCS
bounce in the presence of inhomogeneities?   Second, what would be the observational consequences of this mechanism?
How would inhomogeneities responsible for large scale structure be seeded in this model?   Since we have a bounce, we
should be able to employ the techniques and physical picture in close semblance to the Ekpyrotic scenario.  A plausible
structure generating scenario might arise from inhomogeneous excitations of the condensate something we plan to
investigate in the future.  A promising route would be to employ the techniques developed in~\cite{ABB} to propagate
perturbations (possibly generated by an ekpyrotic scalar field) across the gap-mediated  bounce as the gap only depends
on the overall volume of the universe and is not affected by fluctuations in the metric, a pre-requisite for the success
of the mechanism advocated in~\cite{ABB}.


\end{document}